\documentclass[onecolumn]{aa}
\usepackage{color}
\usepackage{graphicx}
\usepackage{amssymb}
\usepackage{amsmath}
\usepackage{epsfig}

\usepackage{txfonts}
\begin{document}
  \title{
    Indirect Search for Dark Matter in M31 
    with the CELESTE Experiment}
  
  \author{
    J.~Lavalle\inst{1}\fnmsep\thanks{
      \emph{Present address:} 
      Centre de Physique des Particules de Marseille, 
      CNRS-IN2P3 / Universit\'e Aix-Marseille II, 13288 France}
    \and H. Manseri\inst{2}
    \and A. Jacholkowska\inst{1}
    \and E. Brion\inst{3}
    \and R. Britto\inst{1}
    \and P. Bruel\inst{2}
    \and J. Bussons Gordo\inst{1}	 
    \and D. Dumora\inst{3}
    \and E. Durand\inst{3}
    \and E. Giraud\inst{1}
    \and B. Lott\inst{3}
    \and F. M\"unz\inst{4}
    \and E. Nuss\inst{1}
    \and F. Piron\inst{1}
    \and T. Reposeur\inst{3}
    \and D.A. Smith\inst{3}}

  \institute{Laboratoire de Physique Th\'eorique et Astroparticules, 
    CNRS-IN2P3 / Universit\'e de Montpellier II, 34095 Montpellier, France
    \and Laboratoire Louis Leprince-Ringuet, 
    CNRS-IN2P3 / \'Ecole Polytechnique, 91128 Palaiseau, France
    \and Centre d'Etudes Nucl\'eaires de Bordeaux-Gradignan, 
    CNRS-IN2P3, 33175 Bordeaux, France
    \and Laboratoire d'Astroparticule et Cosmologie, 
    CNRS-IN2P3 / Coll\`ege de France, 75231 Paris, France}

  \offprints{\tt lavalle@in2p3.fr}
	      
  \date{Received xx, xx; accepted xx, xx}

  \abstract
      {If dark matter is made of neutralinos, annihilation of such Majorana 
	particles should produce high energy cosmic rays, especially 
	in galaxy halo high density regions like galaxy centres.} 
      {M31 (Andromeda) is our nearest neighbour spiral galaxy, and both its 
	high mass and its low distance make it a source of interest for the 
	indirect search for dark matter through $\gamma$-ray detection.} 
      {The ground based atmospheric Cherenkov telescope CELESTE observed M31 
	from 2001 to 2003, in the mostly unexplored energy range 50-500 GeV.}
      {These observations provide an upper limit on the flux above 50 GeV 
	around $10^{-10}\rm{cm}^{-2}\rm{s}^{-1}$ in the frame of 
	supersymmetric dark matter, and more generally on any gamma emission 
	from M31.
      }
      
  \maketitle
  %

  \section{Introduction}\label{intro}
  The presence of dark matter in the Universe has been known for decades. 
  Since early measurements in galaxy clusters (Zwicky,~\cite{zwicky}), the 
  mass distribution of the Universe has been studied at different scales 
  with a focus on dynamical effects by means of galaxy star rotation 
  curves, large-scale galaxy cluster dynamics 
  (Ostriker et al.,~\cite{ostriker}). Recent developments in observational 
  techniques in cosmology have resulted in independent estimates of the matter 
  content of the Universe 
  $\Omega_{\rm{m}}{h^{2}}$, through large-scale structure surveys 
  (Hawkins et al.,~\cite{2dF} ; Loveday et al.,~\cite{sdss}) and 
  measurements of the cosmic microwave background radiation (the most 
  recent being the WMAP mission (Spergel et al.,~\cite{wmap})). Given standard 
  cosmology, all suggest that most of the matter in the Universe is dark, 
  cold and non-baryonic. Such hypotheses have led to the construction of the 
  Cold Dark Matter (CDM) paradigm (Blumenthal et al.,~\cite{blumenthal}): 
  dark matter would be made of Weakly Interacting Massive Particles (WIMPs) 
  which are neutral, stable and originating from the Big Bang era 
  (Lee \& Weinberg,~\cite{lee}).
  
  Supersymmetric theories (SUSY) (see for instance Nilles,~\cite{nilles}) 
  offer an excellent WIMP candidate (Goldberg,~\cite{goldberg}), the 
  neutralino, which is a mixture of the superpartners of the neutral Higgs 
  bosons and of the electroweak gauge bosons. We will not discuss the case 
  of extra-dimension phenomenological theories, which also provide 
  interesting candidates (Servant \& Tait,~\cite{servant}).

  The nature of dark matter is presently probed in both direct searches, by 
  means of underground experiments that could detect elastic interactions of 
  neutralinos with nuclei, and indirect searches using ground based 
  or satellite telescopes to detect cosmic rays (gamma, leptons or hadrons) 
  created by neutralino pair annihilations in galactic or extragalactic media 
  (for review, see Bergstr\"om~\cite{bergstrom2}). 

  These different types of searches, together with collider 
  experiments, are necessary to constrain in a wider view the quantum 
  nature of dark matter, because they allow either additional 
  or complementary surveys of the particle model parameter spaces.

  Searching for WIMP annihilation signatures with ground based $\gamma$-ray 
  telescopes leads to the question of the choice of targets. A good candidate 
  will have a large amount of dark matter, and combine as big a density as 
  possible, as small a distance from us as possible, and finally, will transit 
  at high elevation in the experimental sky. The Galactic centre is a prime 
  candidate except for being too close to the horizon for CELESTE, which is 
  in the northern hemisphere. Instead, we have chosen M31 and the Draco dwarf 
  galaxy for our searches. M31, which is the nearest giant spiral galaxy, 
  is very massive ($\sim 10^{12} M_{\odot} $), and its star rotation curve 
  indicates a large amount of dark matter. Draco, a neighbour dwarf 
  spheroidal galaxy dominated by a dark component 
  (Kleyna et al.,~\cite{kleyna}), is also a very good candidate but our 
  attempts to study it were foiled by bad weather.

  In this paper, we present the result of searches for $\gamma$-ray emission 
  from M31 with the CELESTE telescope. In section~\ref{predictions}, 
  we review the predictions made for the observations 
  (more details in Falvard~\cite{falvard}, hereafter F04), 
  for which we considered CDM in the frame of minimal supergravity (mSUGRA) 
  phenomenology. The impact of the astrophysical modelling is briefly 
  revisited, as well as possible consequences of non-standard cosmologies. 
  Focusing on the experimental techniques, we present in 
  section~\ref{celeste} the method we 
  use to search for a $\gamma$-ray signal, with explicit 
  comparisons to Crab data. In the absence of a detection, 
  $2\sigma$ confidence level limits are computed for all studied 
  SUSY models.


  \section{Gamma-ray flux predictions for supersymmetric annihilating 
    dark matter in M31}
  \label{predictions}
  Under the assumption that neutralinos have an isotropic and homogeneous 
  velocity distribution, which is likely to be the case in halo centres 
  where the WIMP velocity is expected to be low, the averaged $\gamma$-ray 
  flux due to their annihilation, integrated above an energy threshold 
  $E_{\rm{th}}$ and within the solid angle $\Delta\Omega$ can be written as:
  \begin{equation}
    \label{flux}
    \Phi(E_{\rm{th}}) = \frac{1}{4\pi} 
    \frac{ N_{\gamma}(E_{\rm{th}})< \sigma v> }{2m_{\chi_0}^2} 
    \int_{\Delta \Omega(\theta)}\int_{\rm{l.o.s.}} \rho^2(s) ds\;d\Omega
    \equiv \frac{1}{4\pi} 
    \frac{N_{\gamma}(E_{\rm{th}})<\sigma v>}{2m_{\chi_0}^2} \;\Sigma(\theta)
  \end{equation}
  We thus decouple astrophysics modelling from SUSY contributions. The first 
  part of the right hand term is related to particle physics, via the 
  thermally averaged product of the cross section $\sigma$ with the velocity 
  $v$ producing $N_{\gamma}(E_{\rm{th}})$ photons of energy $E>E_{\rm{th}}$, 
  and the neutralino mass $m_{\chi_0}$. The second term refers to the 
  (squared) halo density profile $\rho$ integrated within an experimental 
  field of view of angular radius $\theta$ along the line of sight $ds$. At 
  the same time, we define $\Sigma(\theta)$ as the astrophysical factor of 
  the flux.

  As $\gamma$-rays result mainly from hadronization of annihilation final 
  states (mainly quarks and gauge bosons), their spectral shape mainly takes 
  its origin in the decay of $\pi^0$. It has been shown by several authors 
  (see for instance Bergstr\"om et al.,~\cite{bergstrom}, or Tasitsiomi et 
  al.,~\cite{tasitsiomi}) that such a spectrum can be fitted or modelled 
  with respect to the neutralino mass. Therefore, the differential spectrum 
  above a threshold energy $E_{\rm{th}}$ can be written as follows:
  \begin{equation}
    \label{diffspectrum}
    \frac{d\Phi}{dE}(E>E_{\rm{th}}) \equiv 
    \Phi(E_{\rm{th}})\times f(E,m_{\chi_0})
  \end{equation}
  where $f(E,m_{\chi_0})$ is the spectral shape derived from the SUSY model and
  normalized such that $\int_{E_{\rm{th}}}^\infty f(E,m_{\chi_0})dE = 1$, 
  so that $\Phi(E_{\rm{th}})$ is the integrated spectrum above an energy 
  threshold $E_{\rm{th}}$. This expression will be useful when we interpret 
  the M31 data collected by CELESTE.

  
  \subsection{Halo modelling}\label{halo}

  M31 is a late-type Sb spiral galaxy, which lies at a distance of about 
  675 kpc, and is observable from the Northern hemisphere 
  (RA = $10.68^{\rm{o}}$, DEC = $41.27^{\rm{o}}$). A study by Braun 
  (\cite{braun}), based upon the analysis of HI data and a model-independent 
  reconstruction of the velocity field, showed that the star rotation 
  curve arises naturally by considering two optically traced mass components: 
  a bulge, with a total mass of $(7.8\pm0.5)\times 10^{10}\text{M}_{\odot}$, 
  and a disk of $(1.22\pm0.05)\times10^{11}\text{M}_{\odot}$ within 28 kpc. 
  Nevertheless, it seems that the star mass-to-light ratios used in this 
  paper, $\Upsilon_{bulge}= 6.5$ and $\Upsilon_{disk}= 6.4$ (solar units 
  in blue band), are over-estimated (\emph{cf.} F04).

  By lowering the bulge and disk contributions, that is 
  $\Upsilon_{bulge}= 3.7$ and $\Upsilon_{disk}= 2.5$ as indicated by F04, we 
  assumed that a dark halo significantly accounts for the gravitational 
  potential. Let us consider the following CDM density profile:
  \begin{equation}
    \rho_{\text{CDM}}(r) = \rho_0 
    \left(\frac{r_0}{r}\right)^{\gamma}
    \left(\frac{r_{0}^{\alpha}+a^{\alpha}}{r^{\alpha}+
      a^{\alpha}}\right)^{\epsilon},
  \end{equation}
  where $r_0$ usually stands for a core radius and $a$ a scale length. An 
  Navarro-Frenk-White profile (Navarro et al.,~\cite{navarro}), 
  \textit{i.e.} with $\gamma = 1$, $\alpha = 1$ and $\epsilon = 2$, fits 
  the rotation curve, and has its parameters entirely determined by the 
  previous mass-to-light ratios. The fitted values are 
    $\rho_0\sim 0.07\rm{GeV.cm}^{-3}$, $r_0\sim 20\rm{kpc}$ and 
    $a\sim 5\rm{kpc}$. The resulting contribution to the $\gamma$-flux 
  within CELESTE's field of view ($\theta = 5\text{mrad}\sim 0.3^{\rm{o}}$, 
  corresponding to a $3.5$~kpc radius) has been calculated:
  \begin{equation}
    \label{sigma}
    \Sigma(\theta = 5\text{mrad}) = 3\times 10^{19}\text{GeV}^2\text{cm}^{-5}.
  \end{equation}
  This result depends strongly on the central tail of the dark halo, 
  but is rather conservative since calculated with a $r^{-1}$ profile.


  \subsection{Probing the SUSY parameter space}\label{susy}
  We choose the minimal supergravity framework (mSUGRA) to scan over the SUSY 
  parameter space. In this frame, a SUSY model can be defined at the 
  unification scale with 5 parameters: the unified scalar mass $m_0$, 
  the unified gaugino mass $m_{1/2}$, the Higgs vacuum expected value 
  ratio $\tan(\beta)$, the unified trilinear coupling $A_0$ and 
  the sign of the mixing parameter of the Higgs superfields $\mu$. 
  We use an interface between the public codes Suspect 
  (Djouadi et al.,~\cite{suspect}) and DarkSusy (Gondolo et 
  al.,~\cite{darksusy}) to compute SUSY masses, annihilation rates and 
  relic densities for various random models. The constraints on these models 
  come from standard accelerator limits, and we select a rather large range 
  for the relic density ($\Omega_{\chi_0}h^2\in[0.05,0.14]$)\footnote{The 
    upper limit is given by the WMAP result plus three sigma. Higher values 
    for the relic density are not that interesting because they correspond to 
    lower values of the annihilation cross-section.}, according to the WMAP 
  measurement $\Omega_{\chi_0}h^2 = 0.113\pm 0.009$ (Bennett et 
  al.,~\cite{bennett}).\\

\begin{figure}
    \centering
    \psfig{figure=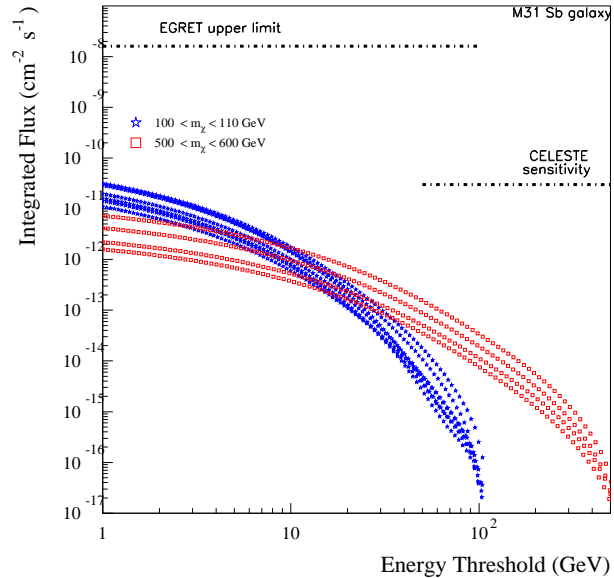,width=9cm}
    \caption{Integrated expected $\gamma$-flux from M31 as a function of the 
      energy threshold, for two selections of low (100-110 GeV) and high 
      (500-600 GeV) neutralino masses ($\Omega_{\chi_0} h^2\in[0.086,0.14]$), 
      showing the mass-dependence of the flux, and the complementarity between 
      EGRET and CELESTE.}
    \label{fig1}
  \end{figure}

  Combining the resulting $\gamma$-spectrum with the astrophysical 
  factor $\Sigma$, we calculate the integrated flux as a function of the 
  threshold energy, as shown in Fig.~\ref{fig1}, in which we have selected 
  only two groups of WMAP compatible neutralinos (respectively small and high 
  masses) in order to exhibit the mass-dependence of the expected flux. 
  The SUSY models plotted there are characterized by a high value of 
  $\tan\beta$ (typically $>30$), for which the production of $\gamma$-rays is 
  more efficient (due to high branching ratio in b quarks). On the same plot, 
  we represent the upper limit provided by the EGRET collaboration (Blom et 
  al.,~\cite{blom}) and CELESTE's sensitivity, 
  $\sim 6\times 10^{-11}\rm{ph.cm}^{-2}\rm{s}^{-1}$, estimated for 50 hours of 
  observation. This figure illustrates the mass-and/or-energy experimental 
  complementarities.

  \subsection{Other contributions}
  \label{xtrafactors}
  Although the predicted fluxes are low 
  ($\sim 10^{-13}\text{cm}^{-2}\text{s}^{-1}$ at a threshold of $\sim50$ GeV), 
  $\sim3$ orders of magnitude smaller than for the Crab nebula (therefore 
  far from CELESTE's sensitivity), several effects could enhance them.\\

  First of all, dark matter substructures, the so-called clumps, arise 
  naturally in the hierarchical scheme of galaxy formation, and simulations of 
  the non-linear regime of collapse allow a semi-theoretical study of their 
  statistics and structure (Moore et al.,~\cite{moore}). 
  Such local overdensities should induce extra neutralino annihilations, 
  and translate to an additional factor to the flux. 
  Although it is rather difficult to estimate how clumpy the galaxies 
  remain today, this enhancement factor was until recently supposed to be 
  smaller than $\sim 10$ (Stoehr et al.,~\cite{stoehr}). Nevertheless, a 
  recent study by Diemand et al. (\cite{diemand}) suggests that about 50\% 
  of a Milky-Way-like galaxy mass is possibly bound to dark matter 
  substructures, whose mass range spreads from $10^{-6}$ up to 
  $10^{7} M_{\odot}$. The authors claim that about $\sim 10^{15}$ of such 
  substructures may have survived against gravitational disruption, 
  leading to a boost factor of over two orders of magnitude compared to the 
  smooth contribution. M31 being very similar to our galaxy, this statement 
  should also stand for that source.

  As another possible astrophysical effect, the supermassive black hole at 
  the centre of M31 could raise the central halo profile up due to adiabatic 
  accretion (Gondolo \& Silk,~\cite{gondolosilk}). There are other interesting 
  mechanisms involving baryons to enhance the dark matter density. Gnedin et 
  al. (\cite{gnedin}) sketch such responses of halos to condensation of 
  baryons, while Bertone et al. (\cite{bertone}) propose a possible 
  enhancement of the WIMP annihilation rate due to the presence of 
  intermediate-mass black holes. \\

  Beside those astrophysical effects, some recent developments in the 
  frame of theoretical cosmology have focused on the quintessence scheme 
  (Caldwell et al.,~\cite{caldwell}) to solve the so-called coincidence 
  problem (the fact that $\Omega_{\Lambda}\sim\Omega_{\text{matter}}$ today). 
  Such a quintessential field could undergo a \textit{kination} regime in 
  the early universe (Salati,~\cite{salati}), so that its kinetic energy 
  dominates over its potential. In this regime, the expansion rate of the 
  universe is enhanced and the thermal history of neutralinos is consequently 
  modified: the decoupling of neutralinos can take place more rapidly at 
  earlier times. Therefore, the WMAP constraint leads to a higher neutralino 
  annihilation cross-section. This means that this phenomenon rehabilitates 
  SUSY models for which relic densities are too low, when calculated in 
  standard cosmology. Salati (\cite{salati}) shows that the relic density 
  enhancement can be parameterized by:
  \begin{equation}
    \label{eta}
    \Omega_{\chi_0}\rightarrow\tilde{\Omega}_{\chi_0} 
    \simeq 1000\left(\frac{m_{\chi_0}}{100\text{GeV}}\right)
    \sqrt{\eta_0}\,\Omega_{\chi_0}\;\text{with}\;\eta_0\leq 0.3.
  \end{equation}
  $\eta_0\equiv\rho_{\Phi,0}/\rho_{\gamma,0}$, where the 0-index refers to 
  a temperature of 1 MeV, and $\rho_{\Phi}$ (respectively $\rho_{\gamma}$) 
  is the quintessence (photon) energy density. The upper limit on $\eta_0$ 
  comes from Big Bang Nucleosynthesis stages that should not be perturbed by 
  the kination regime (Yahiro et al.,~\cite{yahiro}).
  
  According to this cosmology, higher annihilation rate models are required, 
  which therefore means that the $\gamma$-ray production is enhanced. This 
  effect appears in the final results showed in Fig.~\ref{fig7}, 
  for a small sample of selected SUSY models.\\

  Finally, we emphasize another interesting effect coming from Affleck-Dine 
  baryogenesis in SUSY (Fujii et al.,~\cite{fujii}), which yields natural 
  matter-antimatter asymmetry in the early universe. In such a scenario, 
  meta-stable particles result from oscillations in flat directions of the 
  scalar potential, carrying baryon and/or lepton number, namely Q-balls. 
  These Q-balls can have a lifetime long enough to decay after the freeze-out 
  of neutralinos. This induces a non-thermal production of neutralinos, 
  and thus enhances their relic density. This also requires, as previously, 
  higher neutralino annihilation rates to not overclose the universe.\\

  Therefore, although standard conservative predictions are not that 
  optimistic, all these putative contributions may increase the $\gamma$-flux 
  from M31 significantly. This further motivates observations of such a source 
  with CELESTE, keeping in mind that CDM could be something besides SUSY.


  \section{Observations of M31 with CELESTE}\label{celeste}


  \subsection{The CELESTE experiment}\label{setup}

  CELESTE (Par\'e et al.~\cite{pare}, de Naurois et al.~\cite{denaurois}) is 
  an atmospheric Cherenkov telescope detecting $\gamma$-rays above 
  $\sim50$~GeV (the experiment shut down in June 2004). 
  Reaching such a low energy threshold required a large light collection 
  area, achieved by exploiting the mirrors of a solar plant. 
  These mirrors are used to sample the arrival time 
  and photon flux of the Cherenkov wavefront generated by atmospheric showers 
  initiated by cosmic rays at many points in the light pool. In contrast to 
  the imaging technique (Weekes,~\cite{weekes}) the sampling technique uses 
  information on the shape of the wavefront for hadron rejection, as described 
  below.

  The CELESTE experiment uses 53 heliostats (40 until 2001) of the Th\'emis 
  former solar plant (French Pyr\'en\'ees, $42.50^{\rm{o}}$N, 
  $1.97^{\rm{o}}$E, altitude 1650~m). Each heliostat (54 $\text{m}^2$) 
  reflects the light onto the secondary optics, located at the top of a 100 
  meter tower, focussing the light onto a single photomultiplier (PMT) for 
  each heliostat. The PMT signals are sent to the trigger electronics and to 
  the data acquisition system where they are digitized by $\sim$~1~GHz flash 
  analog-to-digital converters (FADCs).

  The mean altitude of the maximum Cherenkov emission for $\gamma$-ray 
  induced showers is around 11~km above the site. The heliostats are aimed at 
  this altitude in the direction of the source under study to enhance light 
  collection. The observations are made in the On-Off tracking mode: the 
  observation of the source (On) is followed or preceded by an observation 
  at the same declination offset in right ascension by 20 minutes. The latter 
  is used as a reference for the cosmic-ray background and the signal is 
  given by the difference between On and Off, after analysis cuts.

  M31 is a special source for CELESTE in that its blue magnitude is about 4.3, 
  and $\sim 5.4$ if integrated in the $\pm 5$ mrad field of view of CELESTE 
  (de Vaucouleurs,~\cite{devaucouleurs}). Hence, pointing On-source increases 
  the PMT illumination compared to the Off-source data. The same problem 
  arises to a lesser degree for the study of the blazar Mrk421, due to the 
  presence of the $m_B=6.16$ star 51 U Ma in the same field-of-view. On-off 
  illumination differences introduce biases at the trigger level and in the 
  recorded data which fake a signal if not handled properly. We remove these 
  biases by ``padding'' the Off-source data with extra background light as 
  part of the data analysis, and by making a pulseheight cut 10\% above the 
  hardware trigger threshold, as described in (de Naurois et 
  al.,~\cite{denaurois}) and updated in (Manseri,~\cite{manseri}). 

  In the following, all Monte Carlo simulations are performed 
  at the transit position of sources in the Th\'emis sky, unless 
  specified. We have used the atmospheric shower simulator Corsika 
  (Heck et al.,~\cite{corsika}) for our Monte Carlo studies. Moreover, 
  stellar photometry studies using the PMT anode currents provided an 
  improved description of our optics in the detector simulation 
  (Smith \& Brion,~\cite{smithandbrion}).


  \subsection{Hadron rejection}\label{rejection}

  In our energy range, the Cherenkov wavefront from $\gamma$-ray induced 
  showers is, on average, more spherical than for showers initiated by hadrons
  (that is, by charged cosmic rays, mainly protons). The CELESTE field of view 
  is small compared to the angular extent of the showers, which lessens the 
  difference, but efficient hadron rejection is still possible.

\begin{figure}
    \centering
    \psfig{figure=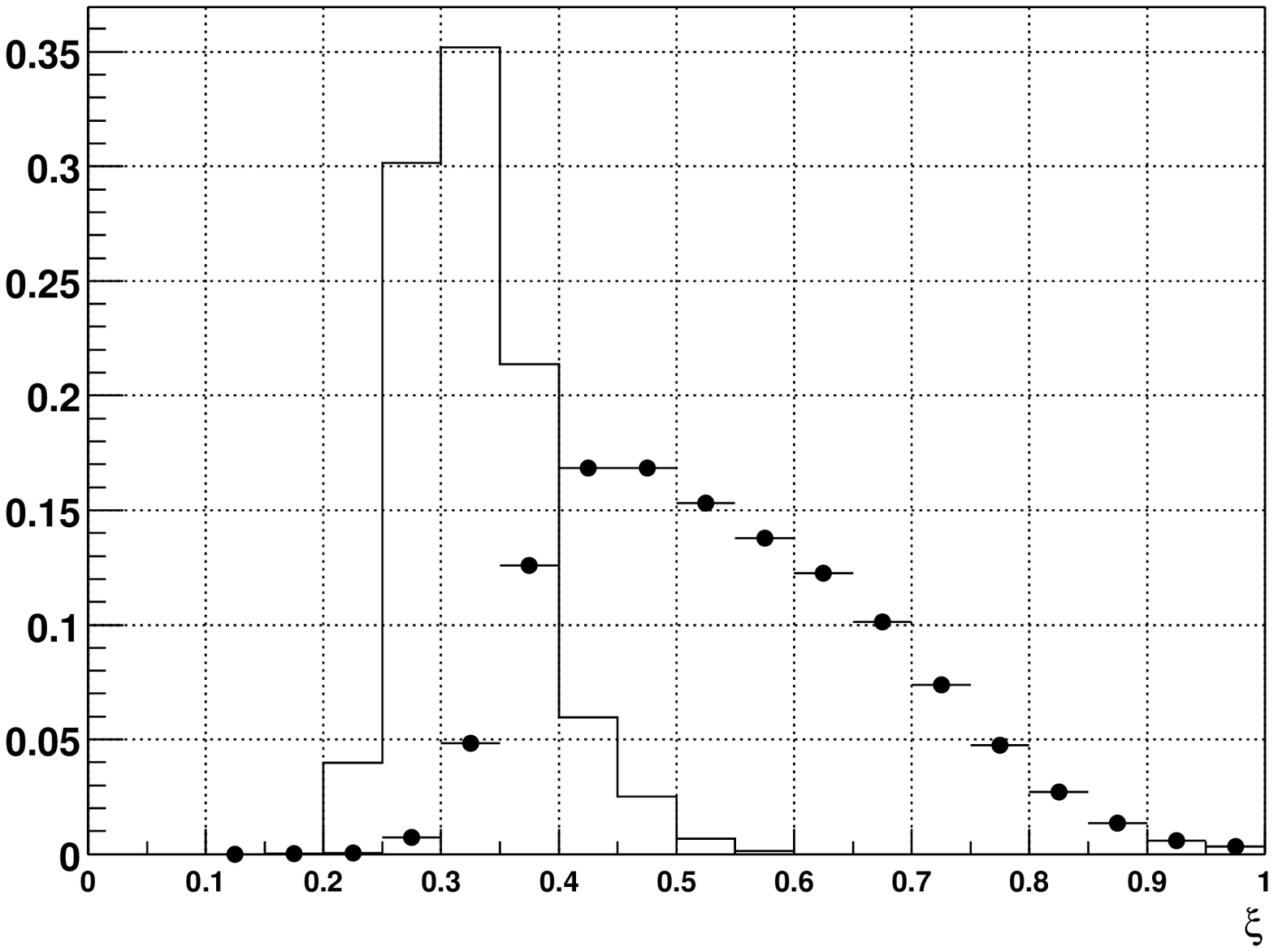,width=8cm}
    \psfig{figure=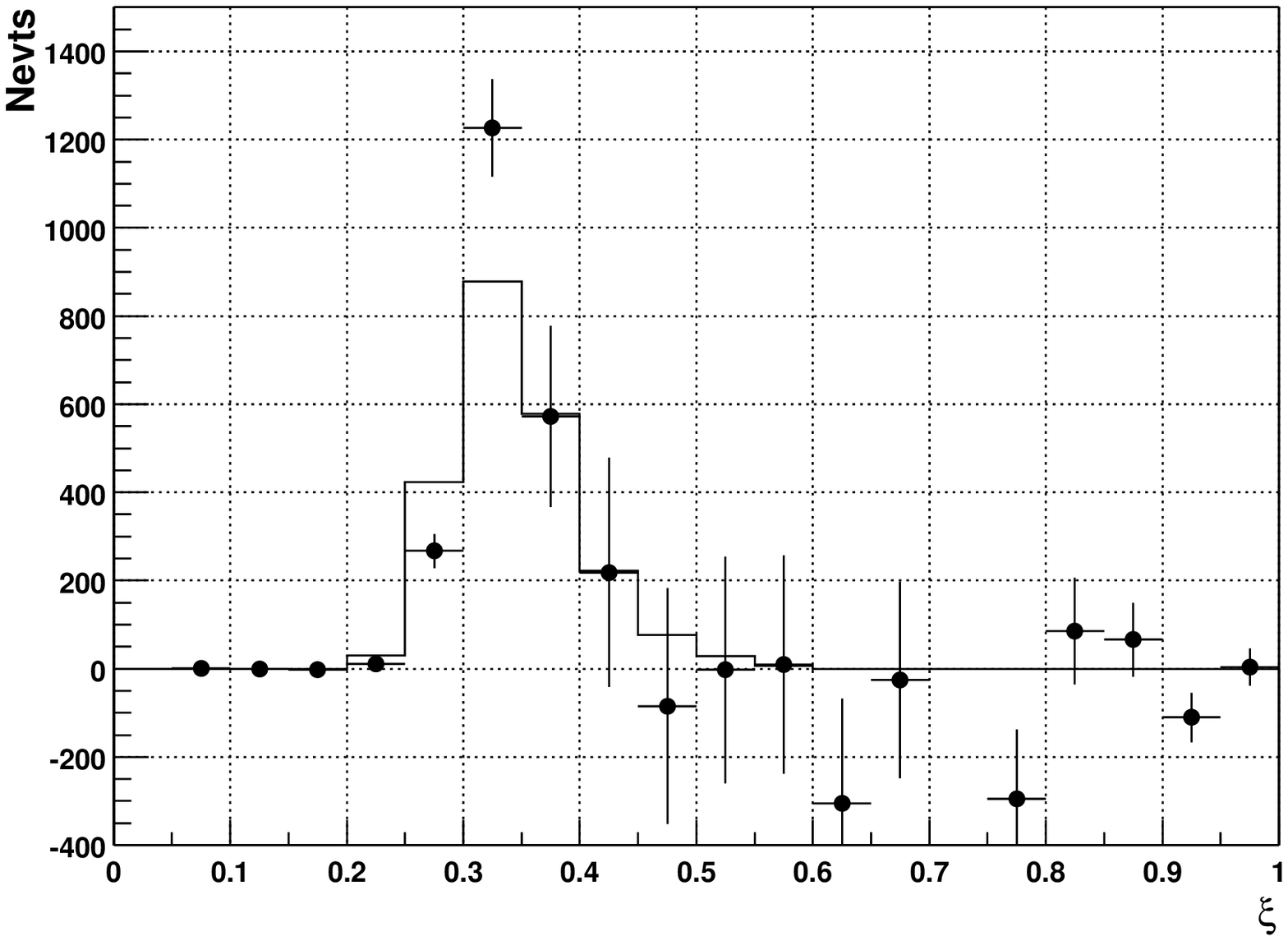,width=8cm}
    \caption{Left: normalized distributions of the discriminating analysis 
      variable $\xi$ for a simulation of $\gamma$-rays (solid line) from M31 
      following a CDM-like spectrum ($m_{\chi_0} = 500$~GeV), with respect to 
      M31 Off-data (2001-03, black markers). Right: number distribution of 
      $\xi$ for a $E^{-2}$-spectrum simulated at the Crab transit (solid 
      line), with respect to the On-Off difference for a data sample of the 
      Crab nebula (taken between 2002 and 2004, black markers with error 
      bars).}
    \label{fig2}
  \end{figure}

  Just above the trigger threshold, the Cherenkov signal for many heliostats 
  is comparable to the fluctuations of the night sky background light, so we 
  sum all the signals instead of using each individually. Summing the signals 
  implies compensating for the propagation delays, which requires knowledge 
  of the shower core position when assuming a spherical Cherenkov wavefront. 
  As described in Manseri~\cite{manseri}, this is done by scanning the plane 
  at 11~km and evaluating the ratio $H/W$ for each position of the scan, 
  where $H$ and $W$ are the height and the width of the summed signal. The 
  largest value of the ratio $H/W$ yields our measure of the shower core 
  position. The sphericity of the wavefront is estimated by how much the 
  ratio $H/W$ decreases when estimated 200~m away from the shower core 
  position. This relative decrease, called $\xi$, is shown in 
  Fig.~\ref{fig2}-left for an Off observation and for a simulation of a 
  $\gamma$-ray spectrum. As expected, because of their Cherenkov wavefront 
  sphericity, $\gamma$-ray showers have lower $\xi$ values than hadronic 
  showers.

  The On-Off difference of this relative decrease is shown in 
  Fig.~\ref{fig2}-right for a sample of Crab nebula data, which was 
  taken between 2002 and 2004 with the same experimental setup as for M31. 
  It exhibits a clear excess due to the $\gamma$-ray emission from the Crab 
  nebula. Requiring $\xi<0.35$ leads to a $13.5\sigma$ excess and a 
  sensitivity of $6.5 \sigma/\sqrt{h}$.

  \subsection{Observations, data selection and signal searches}\label{analysis}

  M31 was observed with CELESTE from 2001 to 2003, and 68 On-Off pairs 
  were collected ($\sim$22 hours of On-source data). Nevertheless, 
  variations in atmospheric conditions are known to cause systematic 
  shifts in the On-Off difference, so we applied a selection based on 
  criteria requiring stable detector operation (characterized by PMT 
  anodic current and trigger rate stability). This selection reduced 
  the data set to $6.5$ hours because of bad weather conditions at 
  Th\'emis since 2001.

\begin{figure}
    \centering
    \psfig{figure=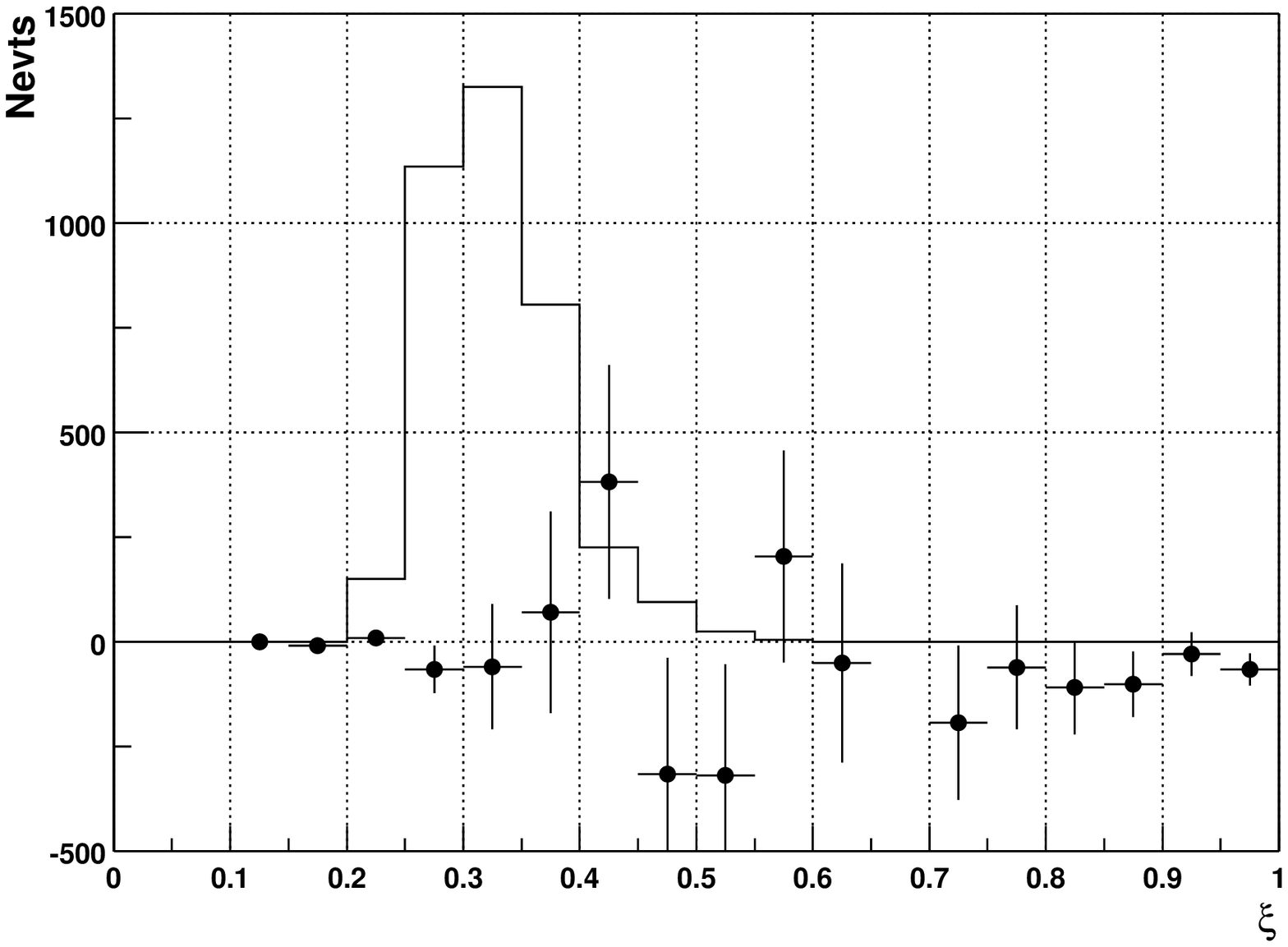,width=8cm}
    \psfig{figure=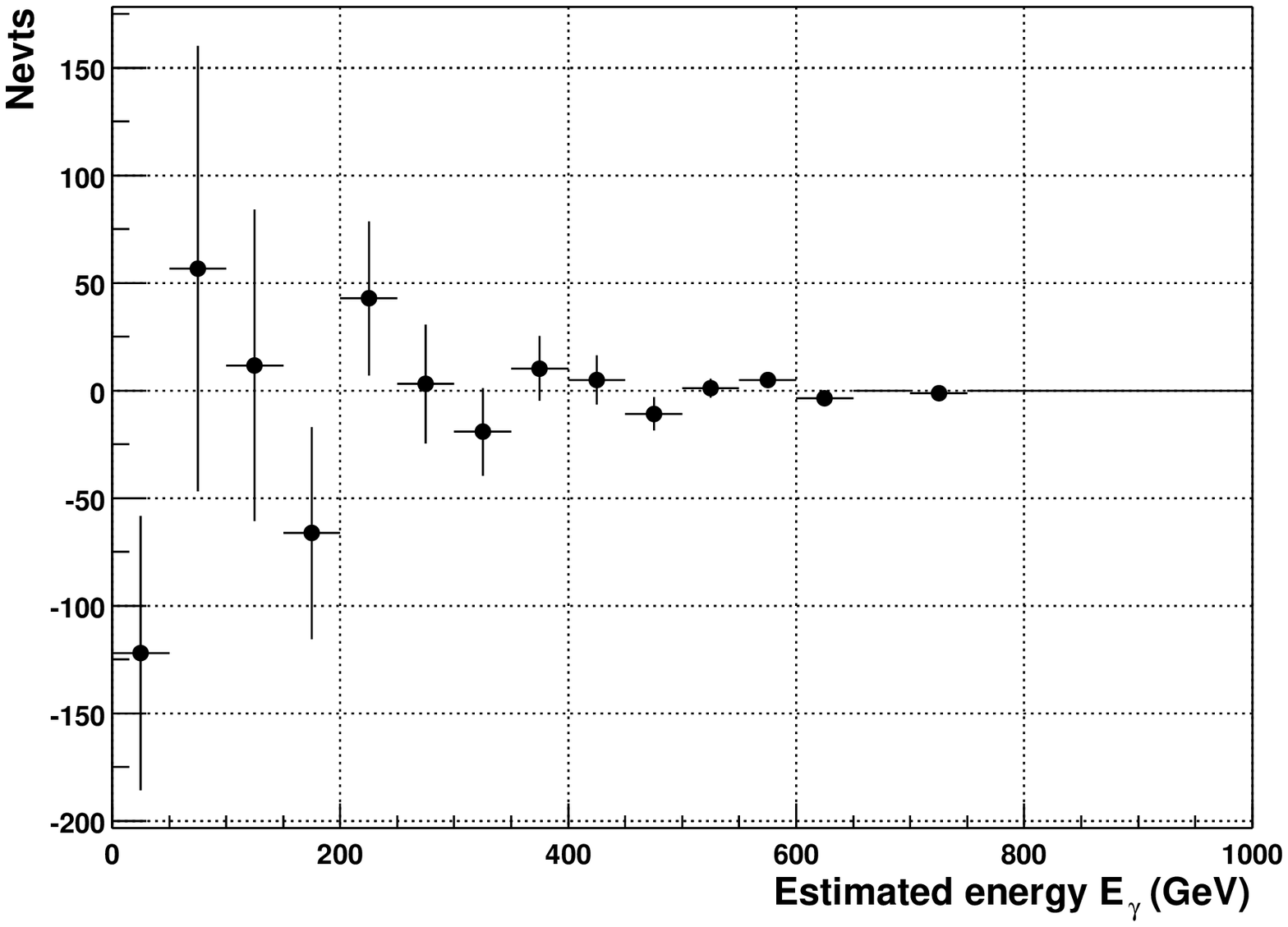,width=8cm}
    \caption{Left: On-Off distribution of $\xi$ for M31 data (2001-03, black 
      markers with error bars), with respect to the same simulated spectrum as 
      Fig.~\ref{fig2}-left (solid line). Right: On-Off distribution of the 
      estimated energy for all events within analysis cuts.}
    \label{fig3}
  \end{figure}

  The On-Off difference of $\xi$ for the M31 data is shown in 
  Fig.~\ref{fig3}-left. No evidence of an excess can be found, and the On-Off 
  difference is $-0.75\sigma$ when requiring $\xi<0.35$.

\begin{figure}
    \centering
    \psfig{figure=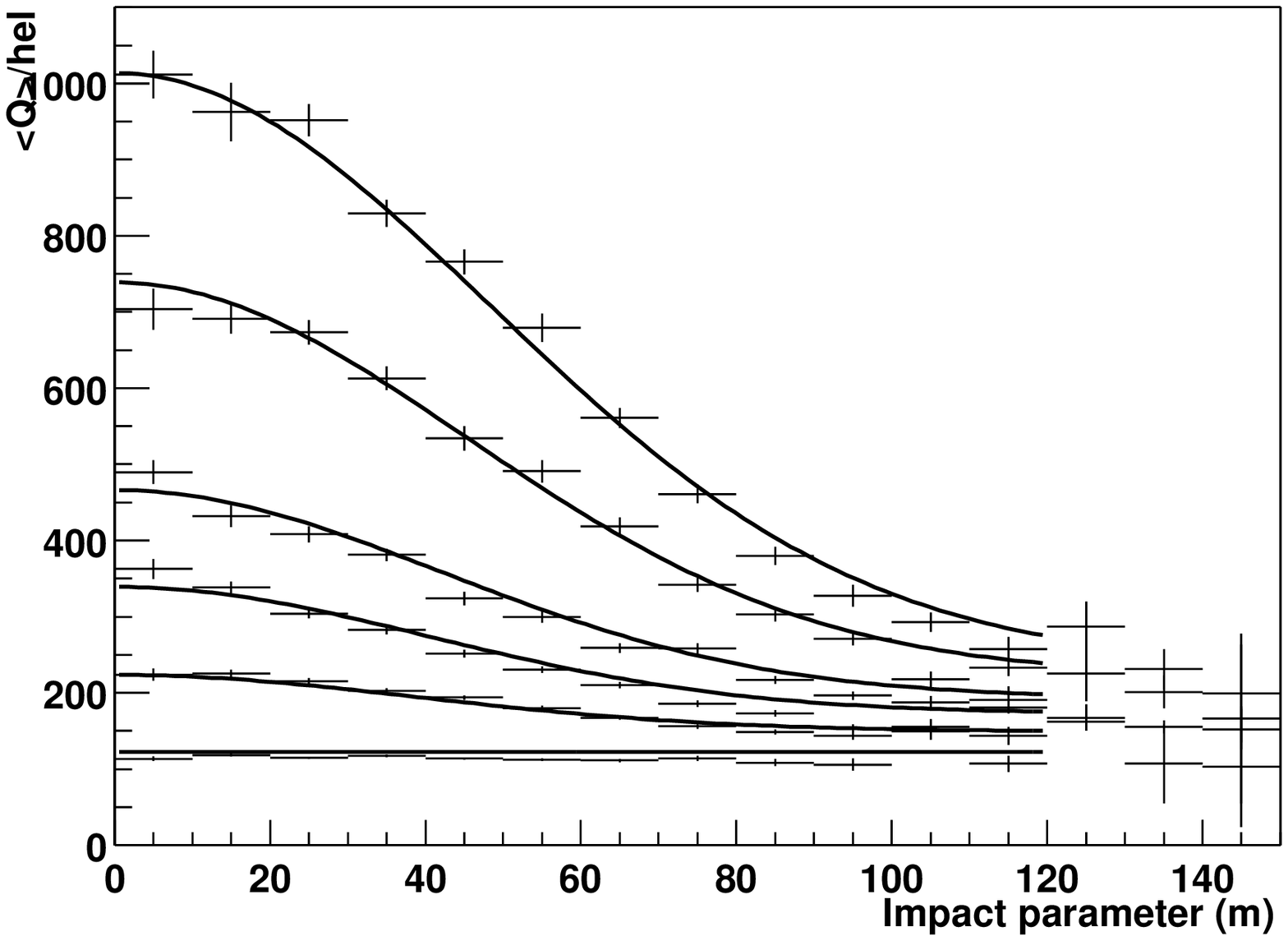,width=8cm}
    \psfig{figure=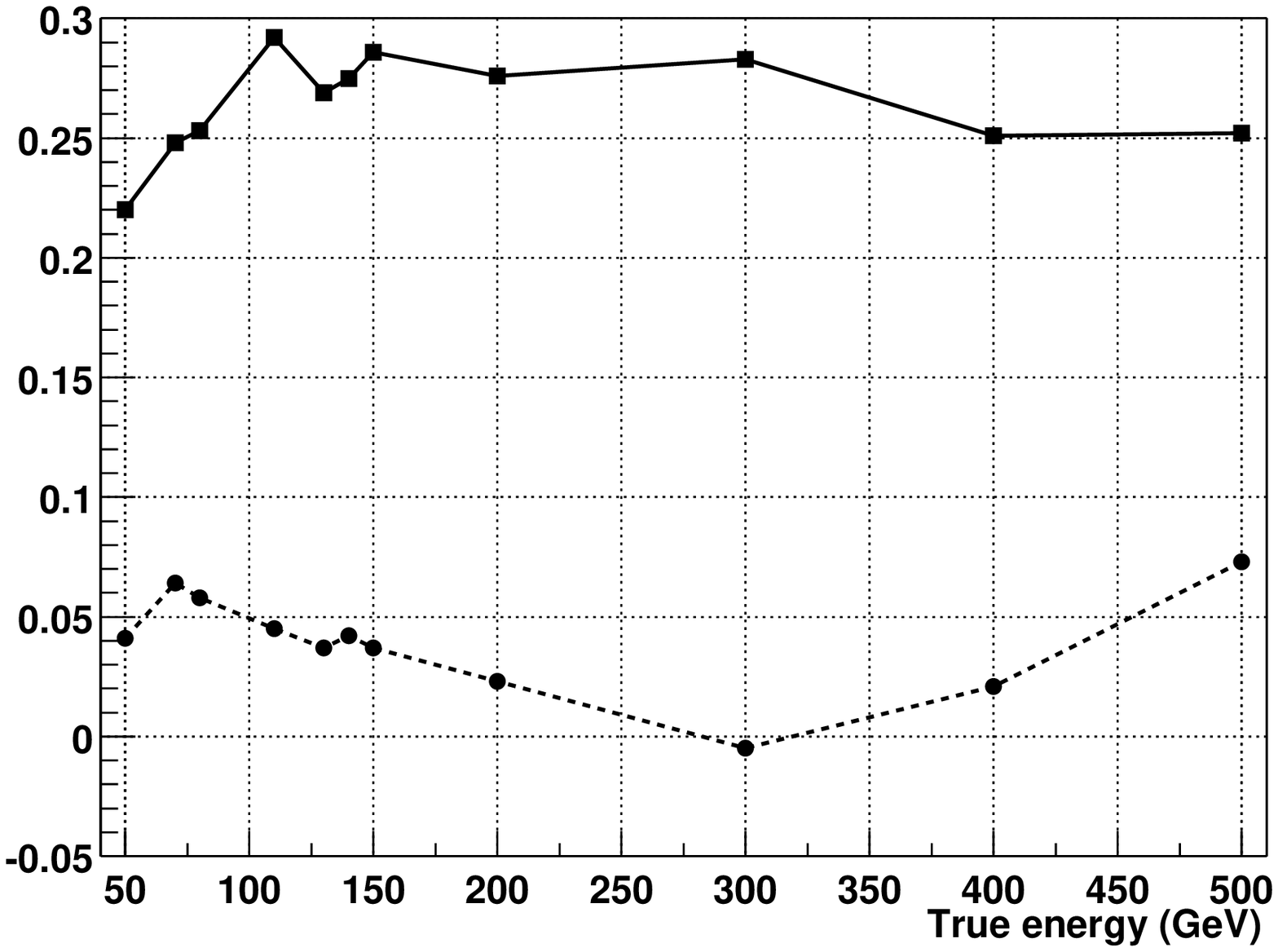,width=8cm}
    \caption{Left: The crosses are the mean charge per heliostat (arbitrary 
      units) for $\gamma$-rays simulated with energies of 50, 100, 150, 200, 
      300 and 400 GeV (bottom-top), versus the reconstructed core position 
      ({\em i.e.} the impact parameter). The solid curves show the charge 
      obtained from the functions used to predict the energy. Right: 
      resolution (solid upper curve) and bias (lower dashed curve) for the 
      energy reconstruction method with respect to the true energy (same 
      Monte-Carlo set-up as in plot at left).}
    \label{fig4}
  \end{figure}

  SUSY annihilating dark matter could produce soft spectra or peaked signals, 
  so we search for a signal in various energy bands. Our energy reconstruction 
  uses $\gamma$-ray simulations with fixed energies, and is based 
  on the correlation between the total charge recorded with the FADCs 
  and the true ({\em i.e.} Monte Carlo) energy, for a given impact parameter. 
  (The shower core position at 11 km, 
  obtained by maximizing $H/W$ as described above, gives the impact parameter 
  on the ground assuming that the gamma ray comes from the source under study.)
  Figure~\ref{fig4} (left) shows the mean charge per heliostat versus 
  the reconstructed impact parameter, for different $\gamma$-ray energies. 
  We use the charge distributions at each energy and impact parameter to 
  build the inverse function, that is, the function predicting the energy 
  from the observed charge and the reconstructed impact parameter in the 
  range of $[0-120]$~m. To check the function, we inverse it again and compare 
  it to the original points, also shown in Fig.~\ref{fig4} (left). The right 
  panel of Fig.~\ref{fig4} shows the energy bias and resolution curves 
  obtained from this estimation method. The energy resolution is below 30\% 
  with a small bias of $\sim$ 5\%, adequate to search for an excess in the 
  data.

  The On-Off difference of the measured energy distribution,  after analysis 
  cuts, is shown in Fig.~\ref{fig3}-right, in which the 50 GeV binning covers 
  at least one standard deviation of the energy resolution function. No excess 
  has been found. All results are summarized in Table~\ref{table1}, in which 
  a search for an excess within bins of 100 GeV, much larger at low energy 
  than the resolution, is presented.

  \begin{table}
    \centering
    \begin{tabular}{|c|c|c|c|c|}\hline
      Analysis level & On-source evts & Off-source evts & 
      On-Off & Significance ($N_{\sigma}$)\\
      \hline
      \hline
      Raw data & 463520 & 462327 & 1193 & -0.59 \\
      \hline
      \hline
      Analysis cuts & 10615 & 10740 & -125 & -0.75 \\
      \hline
      $E^{\rm{meas}}<100$ GeV &  6101 &  6167 &  -66 & -0.53 \\
      \hline
      $100\leq E^{\rm{meas}}<200$ GeV & 3143 & 3197 & -54 & -0.61\\
      \hline
      $200\leq E^{\rm{meas}}<300$ GeV & 870 & 824 & 46 & 1.02 \\
      \hline
      $300\leq E^{\rm{meas}}<400$ GeV & 260 & 269 & -9 & -0.35 \\
      \hline
    \end{tabular}
    \caption{Final statistics resulting from M31 data analysis. The analysis 
      cuts are the following: we impose a software trigger 10\% higher than 
      the hardware one, and we apply a cut $\xi<0.35$ (see text for details) ; 
      for signal searches in energy bins, we add a selection on the 
      reconstructed impact parameter -- $<120$~m -- according to the validity 
      range of our energy reconstruction method. No significant excess appears 
      over the whole sample, nor within different ranges of energy.}
    \label{table1}
  \end{table}


  \subsection{Stability and upper limit}\label{limit}

  \begin{figure}
    \centering
    \psfig{figure=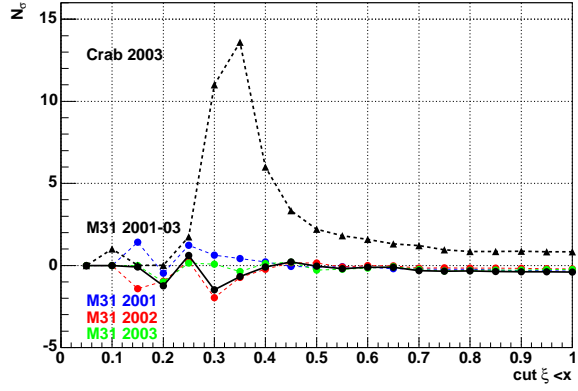,width=8.5cm}
    \caption{Significance level according to varying the cut on $\xi$ (after 
      raising the threshold by 10\%). This figure shows the study of stability 
      of the analysis as well for Crab as for M31 data. A signal evidence 
      is clearly seen for the Crab, whereas the significance level remains 
      flat for M31.}
    \label{fig5}
  \end{figure}

  We have also studied the stability of this result, as a check of possible 
  systematic effects. We show in Fig.~\ref{fig5} how the significance 
  remains stable when varying the cut value on the main discriminating 
  variable $\xi$. As a comparison, the same exercise has been done for a 
  Crab dataset (same figure), and the $\gamma$-ray signal clearly exhibits 
  a bump in significance up to $N_{\sigma}\sim 13.5$ around $\xi<0.35$.

\begin{figure}
    \centering
    \psfig{figure=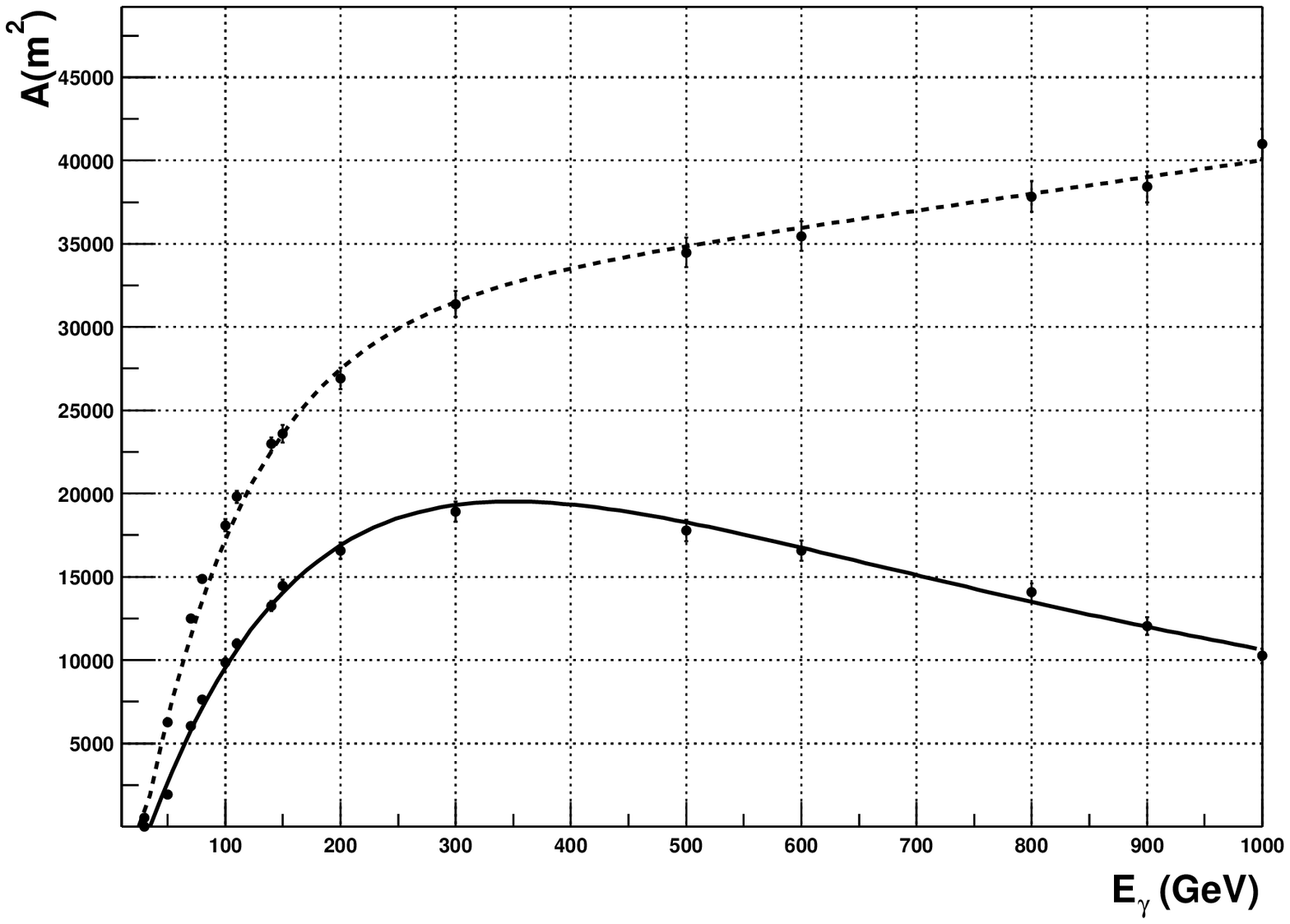,width=8cm}
    \psfig{figure=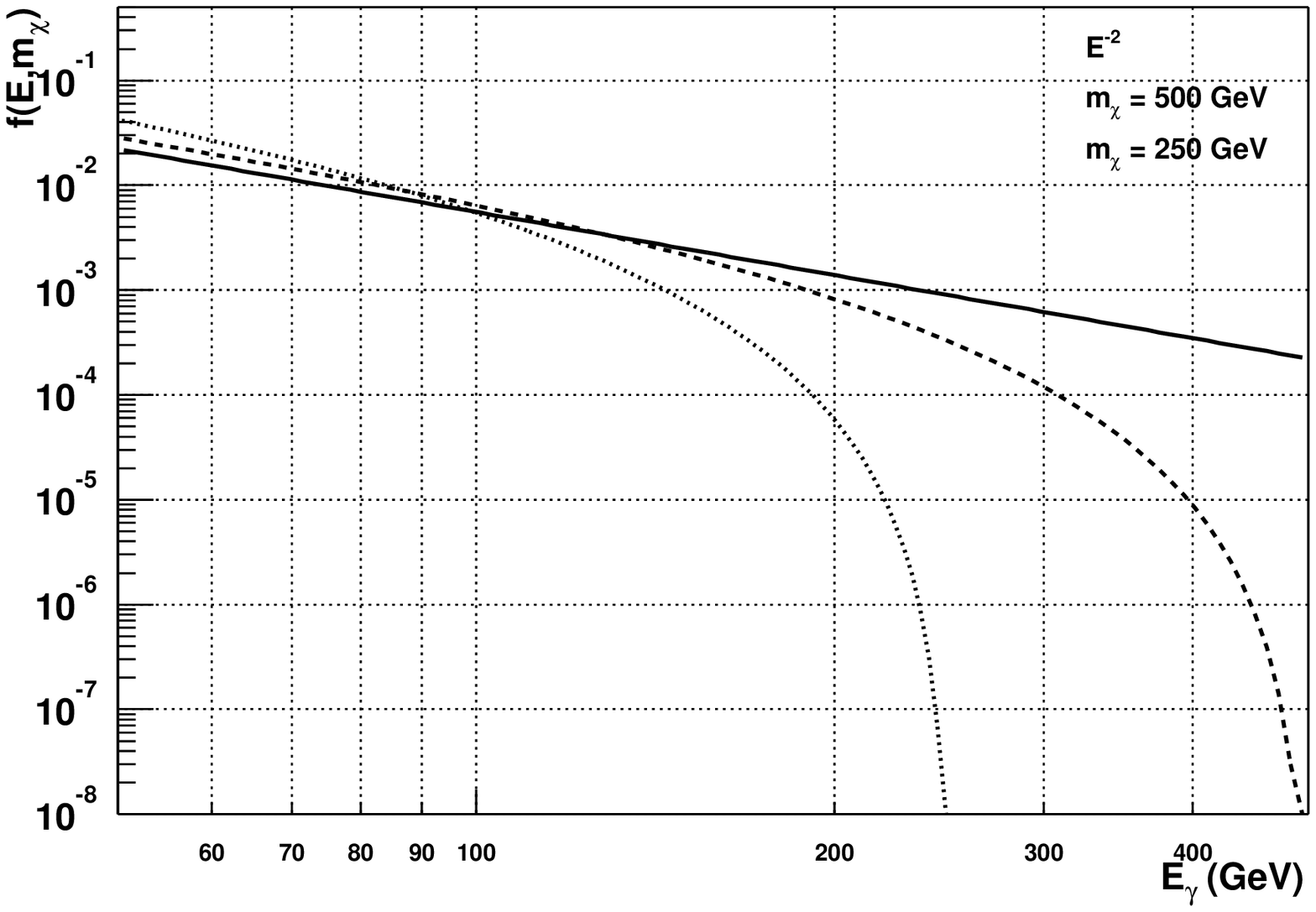,width=8cm}
    \caption{Left: effective detection area -- dashed line for trigger, full 
      after event selection -- as a function of the true energy, 
      based on simulations of $\gamma$-rays coming from the transit position 
      of M31. Right: normalized spectra between 50 and 500 GeV for a power law 
      of index -2, and for formula of equation~\ref{tasitspectrum} 
      with $m_{\chi^0}$ = 500 (dashed curve) and 250 GeV (dotted curve).}
    \label{fig6}
  \end{figure}

  On-Off difference measures the integral over energy of the theoretical 
  spectrum of equation~\ref{diffspectrum} convoluted with the effective area, 
  times the observation time. As the statistics collected is compatible 
  with the absence of a signal, we have only measured the background and its 
  fluctuations in direction of M31. This can be translated to an upper 
  limit on a flux coming from that source, given a normalized theoretical 
  spectral shape $f(E)$. For a $N_{\sigma}$ upper limit, and 
  given an experimental energy threshold $E_{\rm{th}}$, any integrated flux 
  above this energy should be bounded like:
  \begin{equation}
    \label{eqlimit}
    \Phi(E>E_{\rm{th}}) \leq N_{\sigma} \frac{\delta N_{bkgd}}
	{T_{obs}\int_{E_{\rm{th}}}^{\infty}\mathcal{A}(E) f(E) dE}\; 
	\text{(in $\text{cm}^{-2} \text{s}^{-1}$)},
  \end{equation}
  where $\delta N_{bkgd} \simeq \sqrt{2\times N_{Off}}$ is 
  the measured background RMS, $T_{obs}$ is the total exposure time 
  and $\mathcal{A}(E)$ stands for the energy-dependent effective detection 
  area. The latter is determined by means of simulations, and is plotted in 
  Fig.~\ref{fig6}-left. According to equation~\ref{diffspectrum}, note that 
  the spectrum $f(E)$ is mass-dependent in case of neutralino annihilation, 
  so that the limit should depend on the mass. For the sake of simplicity, 
  we will use the parametrization given in Tasitsiomi et al. 
  (\cite{tasitsiomi}) for the spectral shape, that is:
  \begin{equation}
    \label{tasitspectrum}
    f(E) = 
    \frac{6\sqrt{x_{\rm{th}}}}{5 m_{\chi_0}(\sqrt{x_{\rm{th}}}-1)^4}
    \left(
    \frac{10}{3} - \frac{5}{12}\frac{3x^2+6x-1}{x^{3/2}}
    \right)
  \end{equation}
  where 
  $x \equiv E/m_{\chi_0}$ and 
  $x_{\rm{th}}\equiv E_{\rm{th}}/m_{\chi_0}$. This spectrum depends on the 
  mass, but assumes that $\gamma$-rays come only from $\pi^0$ decay. 
  Therefore, it does not take into account all specificities of SUSY models. 
  Nevertheless, it is sufficient for our purpose. This parametrization is 
  illustrated in Fig.~\ref{fig6}-right for two masses, 250 and 500 GeV, and 
  compared with a $1/E^2$ power law spectrum.

\begin{figure}
    \centering
    \includegraphics[width=\textwidth]{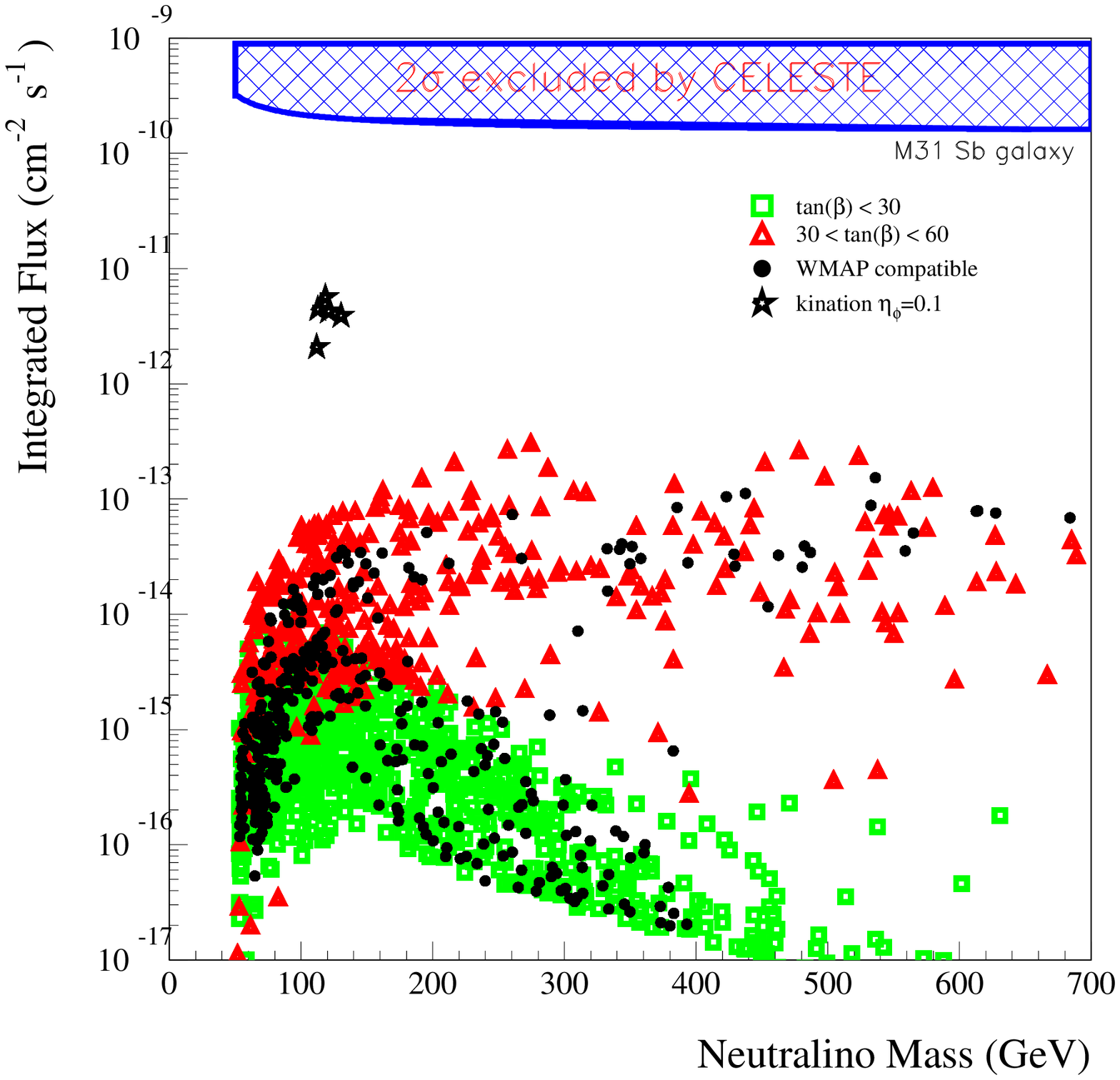}
    \caption{Integrated flux above 50 GeV as a function of the neutralino 
      mass, for SUSY models with $\Omega h^2\in[0.05,0.14]$ (boxes), limited 
      to $[0.086,0.14]$ (full circles) for WMAP compatible models. The 
      $\eta$-parameter is equivalent to the $\eta_0$-parameter discussed 
      in~\S~\ref{xtrafactors}. The dashed region corresponds to the $2\sigma$ 
      CL upper limit from M31 observations with CELESTE.}
    \label{fig7}
  \end{figure}

  The threshold is set to 50 GeV, taken from the effective area shown in 
  Fig.~\ref{fig6} (left). Given this threshold, a flux limit can be computed 
  for each neutralino mass, using equation~\ref{eqlimit}. The result is 
  presented in Fig.~\ref{fig7}, where the predicted integrated fluxes 
  of $\gamma$-rays above 50 GeV are plotted with respect to neutralino masses. 
  The averaged limit in a mass range of $[50-700]$ GeV lies around 
  $10^{-10}\rm{ph.cm.}^{-2}\rm{s}^{-1}$. We emphasize that this is the first 
  experimental result in the energy range 50-500 GeV, and is complementary to 
  those provided by EGRET (Blom et al.,~\cite{blom}) and HEGRA 
  (Aharonian et al.,~\cite{aharonian}) observations of M31.

  \section{Discussion and conclusion}\label{conclusion}
  Observations of M31 with CELESTE provide a $2\sigma$ upper limit on the
  $\gamma$-flux above 50 GeV, depending on the expected spectrum. This limit, 
  around $10^{-10}\rm{ph.cm}^{-2}\rm{s}^{-1}$, is quite far from the SUSY 
  parameter space, but significantly constrains combinations of different 
  enhancement factors discussed in \S~\ref{xtrafactors} (which are also 
  likely to be excluded by EGRET limits, depending on the neutralino mass), 
  and also any other model of annihilating dark matter besides SUSY. 
  Whereas these observations have been motivated by indirect searches for 
  SUSY CDM, this result yields a general astrophysical result: the first 
  observation of a spiral galaxy in this energy range, somehow constraining 
  $\gamma$-ray emission from this class of objects.

  Any gamma ray detection from a galaxy like M31 would be difficult to 
  interpret in terms of dark matter annihilation. Spiral galaxies are known 
  sites of non-thermal processes and cosmic ray acceleration, and the relevant 
  physical mechanisms are not yet well understood. In this sense, a Dwarf 
  Spheroidal galaxy like Draco is a very promising source for indirect 
  detection, given it is clearly dominated by the dark matter component. 
  Unfortunately, we have too few data on Draco to perform a relevant analysis.

  However, the upcoming generation of $\gamma$-ray instruments will 
  undoubtedly further constrain dark matter models and halo models for various 
  astrophysical sources. These searches are not only complementary to future 
  particle physics experiments, but also very important to understand how the 
  question of dark matter is connected to the particle content of the 
  Universe.

  \section*{Aknowledgements}
  Funding was provided by the IN2P3 of the French CNRS and by the 
  Grant Agency of the Czech Republic. We gratefully acknowledge the 
  support of the Regional Council of Languedoc-Roussillon, and of 
  Electricit\'e de France. The technical support by Antoine P\'erez, 
  Jacques Maurand and St\'ephane Rivoire at Th\'emis was invaluable. 
  Last but not least, we appreciatively thank Karsten Jedamzik, Gilbert 
  Moultaka and Pierre Salati for very useful discussions during this work.

\end{document}